\documentclass[amsmath,amssymb,aps,twocolumn,prb,longbibliography,floatfix,superscriptaddress]{revtex4-2}
\usepackage{graphicx,bm,times}
\usepackage[utf8]{inputenc}
\usepackage{braket}
\usepackage{amsfonts}
\usepackage{color}

\renewcommand{\Im}{\operatorname{Im}}

\usepackage[breaklinks]{hyperref}
\hypersetup{
    pdfstartview={FitH},
    colorlinks=true,
    linkcolor=blue,
    citecolor=blue,
    urlcolor=blue
}

\begin{document}

\title{Magnetic Compton profiles of Ni beyond the one-particle picture: \\
numerically exact and perturbative solvers of dynamical mean-field theory}
\author{A.~D.~N.~James}
\affiliation{H.~H.~Wills Physics Laboratory,
University of Bristol, Tyndall Avenue, Bristol, BS8 1TL, United Kingdom}

\author{M.~Sekania}
\affiliation{Institut f\"ur Physik, Martin-Luther Universit\"at Halle-Wittenberg, 06120 Halle/Saale, Germany}
\affiliation{
Theoretical Physics III, Center for Electronic Correlations and
Magnetism,
Institute of Physics, University of Augsburg, 86135 Augsburg, Germany}
\affiliation{Faculty of Natural Sciences and Medicine, Ilia State University, 0162, Tbilisi, Georgia}

\author{S.~B.~Dugdale}
\affiliation{H.~H.~Wills Physics Laboratory,
University of Bristol, Tyndall Avenue, Bristol, BS8 1TL, United Kingdom}

\author{L.~Chioncel}
\affiliation{
Theoretical Physics III, Center for Electronic Correlations and
Magnetism,
Institute of Physics, University of Augsburg, 86135 Augsburg, Germany}
\affiliation{Augsburg Center for Innovative Technologies, University of Augsburg, 86135 Augsburg, Germany}

\date{\today}% It is always \today, today,

\begin{abstract}
We calculated the magnetic Compton profiles (MCPs) of Ni using density functional theory supplemented by electronic correlations treated within dynamical mean-field theory (DMFT). We present comparisons between the theoretical and experimental MCPs. The theoretical MCPs were calculated using the KKR method with the perturbative spin-polarized T-matrix fluctuation exchange approximation DMFT solver, as well as with the full potential linear augmented planewave method with the numerically exact continuous-time quantum Monte Carlo DMFT solver. We show that the total magnetic moment decreases with the intra-atomic Coulomb repulsion $U$, which is also reflected in the corresponding MCPs. The total magnetic moment obtained in experimental measurements can be reproduced by intermediate values of $U$.
The spectral function reveals that the minority X$_2$ Fermi surface pocket shrinks and gets shallower with respect to the density functional theory calculations.
\end{abstract}


\maketitle

%%%%%%%%%%%%%%%%%%%%%%%%%%%%%%%%%%%%%%%%%%%%%%%%%%%%%%%%%%%%%%%%
\section{\label{sec:intro}Introduction}

The electron momentum distribution, through its dependence on the ground state wavefunctions, is a powerful quantity for understanding many-body effects in solids \cite{schneider:92}. Compton scattering experiments, which involve measuring the energy distribution of inelastically scattered photons which have impinged on electrons within the sample being studied, are able to measure a projection (integration over two momentum components) of the underlying electron momentum distribution \cite{cooper:85}. Since the photons scatter from the occupied momentum states, Compton scattering is sensitive to the Fermi surfaces of metals. 

While other experimental techniques such as photoemission spectroscopy (PES, and its angle-resolved counterpart, ARPES) give excellent insight into the many-body interactions present, it is important to remember that the photoemission process is a complex excitation of the whole system. Indeed, any interaction of the photo-hole with the electron quasiparticle would invalidate a claim to be a measurement of the ground-state. Thus Compton scattering is a uniquely powerful probe of the ground state many-body wavefunction \cite{Kaplan_2003}. In recent years, Compton scattering has been used to reveal the electronic structure and Fermi surfaces \cite{dugdale:14} in electronically complex materials such as substitutionally disordered alloys \cite{robarts_2020,billington_2020} and compounds with high vacancy concentrations \cite{ernsting2017vacancies}. Most relevantly, Compton scattering is able to probe the electron correlations within many complex materials  \cite{sakurai:95,schulke:01,ruotsalainen2018isotropic,billington2015magnetic}. Therefore, Compton scattering offers a valuable and complementary perspective on electronic structure and, in particular, a window onto electron correlations in different regimes of composition, temperature and magnetic field from those which other probes can reach.

When X-rays are inelastically scattered by the
electrons in solids, the scattered photon energy
distribution is Doppler broadened because of the
electrons' momentum distribution $\rho(\mathbf{p})$~\cite{willi.77,lo.co.96}. 
In practice this is measured through the double differential scattering cross-section
${\mathrm{d}^2\sigma/ \mathrm{d}\Omega \mathrm{d}\omega}$ for a given infinitesimal solid angle $\mathrm{d}\Omega$ and energy
$\mathrm{d}\omega$ of the scattered photon, respectively. The incident energy of the monochromatic X-rays and the scattering angle are fixed during the experiment, and the scattering cross-section is measured as a function of the photon energy.
If the scattering event is described within the impulse approximation~\cite{ch.wi.52,as.wi.52}, the
scattering cross-section is proportional to the Compton profile, ${\mathrm{d}^2\sigma/\mathrm{d}\Omega \mathrm{d}\omega \propto J(p_z)}$, which is the 1D projection of the electron momentum distribution, $\rho(\mathbf{p})$, along the scattering vector $p_z$:
\begin{equation}
 \label{eq:J}
  J(p_z) = \iint \rho(\mathbf{p}) \mathrm{d}p_x\mathrm{d}p_y\,.
\end{equation}
If the incident photon beam has a component of circular polarization, the scattering
cross-section contains a term which is spin dependent. This term may be isolated from
the charge scattering by either flipping the direction of the sample magnetization or
the photon helicity parallel and antiparallel with respect to the scattering vector, 
resulting in a magnetic Compton profile (MCP), $J_{\mathrm{mag}}(p_z)$ \cite{duffy:13}. 
In analogy to the Compton profile, the MCP is defined as 
the 1D projection of the spin-polarized electron momentum density:
\begin{equation}
 \label{eq:J_mag}
  J_{\mathrm{mag}}(p_z)= \iint 
    \left[
      \rho^\uparrow(\mathbf{p})-\rho^\downarrow(\mathbf{p})
    \right]
    \mathrm{d}p_x \mathrm{d}p_y\,.
\end{equation}

Electronic structure calculations are particularly useful for the interpretation of MCPs and these can be calculated from the spin-dependent momentum distributions. Density functional theory  (DFT)~\cite{ho.ko.64,kohn.99,jo.gu.89,jone.15} is by far the most widely used method with its immense success in predicting properties of the solid state.
However, treating electron correlation in an effective one-particle framework 
results in notable discrepancies with experiment even with the best available functionals. Over the last decade, it has been demonstrated that within the combined DFT and dynamical mean-field
theory (DMFT)~\cite{me.vo.89,ge.ko.96,ko.vo.04}, the so called DFT+DMFT
approach~\cite{held.07,ko.sa.06}, many of the electronic ground state
properties of $d$-metal elements, their alloys and compounds can be
well described~\cite{ko.sa.06,held.07,ka.ir.08}.
Early developments of DFT+DMFT were of the one-shot type of calculation in which a DFT computation is first converged and the subsequent DFT Hamiltonian is supplemented with a local Coulomb interaction for the  correlated orbitals~\cite{sa.ko.04,le.ge.06,held.07,pa.mi.14}. Then, in a separate step, the interacting problem is solved self-consistently within DMFT.
Fully charge self-consistent approaches have been implemented~\cite{ch.vi.03,mi.ch.05,ha.ch.10,ai.po.11} as well, in which both DFT and DMFT are converged simultaneously. For some materials the correlation-induced modification in the charge density can be significant, while in others this was shown not to be the case.
However, to be consistent with the DFT concepts, charge densities ($\rho(\mathbf{r})$), should be potential ($V(\mathbf{r})$) representable (${\rho \rightleftarrows V}$), which is only achieved by full self consistency~\cite{jone.15}.

One of the most studied simple metallic system that presents signatures of electronic 
correlations is the fcc itinerant ferromagnetic Ni. It is known that the DFT alone cannot reproduce the dispersionless feature at a binding energy of about $6$~eV which is known as the ``$6$~eV satellite''~\cite{gu.ba.77}. The valence band photoemission spectrum of Ni shows a $3d$-band width that is about $30\%$ narrower than the value obtained from the DFT 
calculations. Similarly, the exchange splitting in both the local spin-density 
approximation (LSDA) and the generalized gradient approximations (GGA)~\cite{le.ch.91} 
overestimates the experimental splitting by approximately $50\%$ ~\cite{ea.hi.78,di.ge.78,hi.kn.79,ea.hi.80}. 
The combined DFT+DMFT describes the occupied $3d$ bandwidth of Ni, 
reproduces the exchange splitting and the $6$~eV satellite structure in the valence band~\cite{li.ka.01,br.mi.06,sa.br.12}. In addition to this,  DFT+DMFT has shown the consequences of the local moment in ambient and Earth-core-like conditions \cite{hausoel2017local}. Further information about the electronic structure of Ni can be extracted by using Compton scattering.

The MCPs of Ni have been calculated by using various DFT implementations and their extensions. 
Features associated with the Fermi surface (as a consequence of bands crossing the Fermi energy) seen in experiment~\cite{di.du.98} were generally reproduced with good agreement, notwithstanding the distinct discrepancy at low momenta which points towards some inaccuracies in the position of the spin-polarized bands with respect to the Fermi level. 
It has been also shown that the negative polarization of the itinerant $s$- and $p$-like band electrons can be observed~\cite{ku.as.90,ti.br.90} and the discrepancy with respect to the theoretical predictions were attributed to the insufficient treatment of correlations present in the standard DFT exchange-correlation functionals at low momentum~\cite{ku.as.90}. 
The directional Compton and magnetic Compton profiles have also been computed in combination with DMFT ~\cite{be.mi.12,ch.be.14a,ch.be.14b} which facilitated a discussion of the anisotropy of the electronic correlations of Ni as a function of the on-site Coulomb interaction strength, $U$. 
Those theoretical comparisons with the experimental data led to the conclusion that the theoretical MCPs improved when the local correlations are taken into account, which also extends to the total Compton profiles.

In this work, we focus on the calculation of the momentum distribution and related quantities within the framework of many-body theory. 
We have used two approaches on the DFT side, namely KKR~\cite{eber.00,SPR-KKR2.1.1,mi.ch.05} which is a spin-polarized relativistic multiple-scattering theory implementation, and ELK~\cite{elk} which is a full potential linear augmented planewave (FP LAPW) implementation.
For the many-body solvers we used the perturbative SPT-FLEX~\cite{ka.li.02,po.ka.05} in combination with KKR~\cite{mi.ch.05} and the numerically exact continuous-time quantum Monte Carlo (CT-QMC)~\cite{gu.mi.11,seth2016} in combination with the ELK~\cite{elk}.
We show that while the experimental magnetic moment can be obtained by varying $U$, the Ni MCP shapes have a weak dependence on $U$. These results further indicate the importance of the effects beyond the local approximation of DMFT.

%%%%%%%%%%%%%%%%%%%%%%%%%%%%%%%%%%%%%%%%%%%%%%%%%%%%%%%%%%%%%%%%
\section{Compton and Magnetic Compton profiles with DFT+DMFT}
Over the last few decades, substantial progress has been made in the development of computational tools and libraries that combine DMFT with electronic structure methods in the framework of DFT+DMFT~\cite{ko.sa.06,held.07}.
All these implementations can be divided into two sub-groups according to the employed schemes for constructing the local orbitals and the definition of the so-called correlation subspace in which the DMFT equations are solved. One of the sub-groups utilizes the local atomic orbitals constructed from the atomic solutions which cover a wide energy window of the DFT (Kohn-Sham) wavefunctions, and is implemented in almost all existing electron codes with different flavors such as LMTO~\cite{gr.ma.07,gr.ma.12}, EMTO~\cite{ch.vi.03,os.vi.17}, KKR~\cite{mi.ch.05}.
The other popular choice is the set of localized Wannier wavefunctions, which is currently realized as an independent interface in various codes, including VASP~\cite{kr.ha.93}, QUANTUM ESPRESSO~\cite{Giannozzi_2009}, SIESTA~\cite{Soler_2002}, ABINIT~\cite{Gonze2009}, WIEN2k~\cite{bl.sc.20}, and ELK.
The ELK code has been interfaced with the TRIQS/DFTTools application~\cite{james_2021,aichhorn2016} to enable DFT+DMFT calculations with ELK and the TRIQS library \cite{triqs} (we refer to this as the ELK-TRIQS package).
Prior to the description of the computational procedure for the (magnetic) Compton profiles, we briefly present the steps of the fully charge-self consistent DFT+DMFT calculation. 

%%%%%%%%%%%%%%%%%%%%%%%%%%%%%%%%%%%%%%%%%%%%%%%%%%%%%%%%%%%%%%%%
\subsection{DFT(+DMFT) using KKR}
The calculation scheme within the Korringa-Kohn-Rostoker (KKR) method is based on the Green’s function formalism of multiple-scattering theory~\cite{eber.00}. The details of the DFT+DMFT implementation within the fully relativistic multiple-scattering KKR method have been reported previously~\cite{mi.ch.05}. It is important to note that the flexibility of KKR lies in utilizing the Dyson equation to relate the Green's function of a perturbed system to the Green's function of a suitable unperturbed reference system.

In KKR, the central quantity is the multiple scattering path-operator, $\hat{\tau}$, which is used to compute the real-space Green function.
In the most general relativistic form, the scattering path-operator is represented in the site basis with site-index $R$, and the total angular momentum with combined index ${\Lambda(\kappa,\mu)}$, with spin-orbit, $\kappa$, and magnetic, $\mu$, quantum numbers.
Electronic correlations can be included by supplementing the scalar real-valued local single-particle potential provided by the DFT with a self-energy that is energy-dependent, complex-valued, and non-local.
The Dyson equation allows for the computation of the single-particle Green's function ${G({\bf r},{\bf r^\prime},E)}$ with respect to a reference system.
The ${\bf r}$ and ${\bf r^\prime}$ are defined relative to the center of an atomic cell corresponding to a specific site, and the reference system is described by a one-electron Hamiltonian containing the DFT potentials located on the regular lattice sites~\cite{mi.ch.05,mi.eb.17}. 
The single-site Dirac equation is solved and the wavefunction is matched to the free-electron solution at the boundary of the atomic cell. Subsequently, the single-site scattering matrix, ${t^{R}_{\Lambda\Lambda^\prime}(E)}$ is obtained, which in turn provides the expression for the scattering path operator, ${\tau^{R R^\prime}_{\Lambda\Lambda^\prime}(E)}$, connecting the sites $R$ and $R^\prime$. 
Within this basis, employing the four-component wave functions of the regular (${Z^{R(\times)}_\Lambda({\bf r},E)}$) and irregular (${J^{R(\times)}_\Lambda({\bf r},E)}$) solutions of the Dirac equation~\cite{tamu.92}, the Green's function can be written as
\begin{align}
 \begin{split}
    G({\bf r},{\bf r^\prime},E)
    &=
    \sum_{\Lambda \Lambda^\prime} Z^{R}_{\Lambda}({\bf r},E) \tau^{R R^\prime}_{\Lambda\Lambda^\prime}(E) Z^{R^\prime\times}_{\Lambda^\prime}({\bf r^\prime},E)  \\ 
    &- \sum_{\Lambda}\!\left[Z^R_{\Lambda}({\bf r},E)J^{R\times}_{\Lambda}({\bf r^\prime},E)\theta(r^\prime-r) \right. \\ 
    &\quad\ \ + \left.J^R_{\Lambda}({\bf r},E) Z^{R\times}_{\Lambda}({\bf r^\prime},E)\theta(r-r^\prime)\right]\!\delta_{RR^\prime},
 \end{split}
\end{align}
where $\theta(r)$ is the step function.
Note that in the case of a complex non-Hermitian self-energy, ${\Sigma({\bf r},{\bf r^\prime},E)}$, one has to distinguish between the left-and right-hand side solutions of the single-site Dirac equation~\cite{tamu.92}. The left-hand side solutions are denoted by the ``$\times$''-symbol as an upper index.

The DMFT solver used in the current implementation is the relativistic version of the so-called Spin-Polarized T-Matrix (SPT-) Fluctuation Exchange approximation (FLEX)~\cite{ka.li.02,po.ka.05} which is formulated on the complex (Matsubara) axis. 
In contrast to the original formulation of FLEX~\cite{bi.sc.89}, the particle-particle and the particle-hole channels are treated differently~\cite{ka.li.02,po.ka.05}.
The particle-particle processes renormalize the effective interaction, which is added into the particle-hole channel. The particle-hole channel itself describes the interaction of electrons with the spin-fluctuations, which represents one of the most relevant correlation effects in Ni.
Here we employed the SPT-FLEX solver which uses a rotationally invariant formulation for the interaction and is self-consist in charge and self-energy~\cite{mi.ch.05}. Once the self-energy is computed within the many-body solver it is returned directly into the Dirac equation~\cite{mi.ch.05}.

In order to compute the electron momentum density, the momentum operator is diagonalized in the crystal basis set, and its eigenfunctions are used to construct the Green function in the momentum representation, ${G_{\sigma}(\mathbf{p}, E)}$~\cite{be.ma.06}. 
The spin-projected momentum density ${\rho^{\sigma}(\mathbf{p})}$ can then be directly computed as~\cite{be.ma.06}:
\begin{equation}
 \label{eq:rho_KKR}
  \rho^{\sigma}(\mathbf{p}) = {-\frac{1}{\pi} \int\limits_0^{E_F}\Im G_{\sigma}(\mathbf{p},E)dE}\,,
\end{equation}
where ${\sigma=\,\uparrow\!(\downarrow)}$ represents the spin projections. 
The energy integration in Eq.~\eqref{eq:rho_KKR} is performed in the complex plane along the contour that encloses the poles of the one-particle Green's function. The corresponding Compton and magnetic Compton profiles are computed using Eqs.~\eqref{eq:J} and~\eqref{eq:J_mag}.

%%%%%%%%%%%%%%%%%%%%%%%%%%%%%%%%%%%%%%%%%%%%%%%%%%%%%%%%%%%%%%%%
\subsection{DFT(+DMFT) using ELK-TRIQS}

The DFT+DMFT framework within the ELK-TRIQS package (which will be referred to as ELK+DMFT throughout) starts with a self-consistent calculation at the DFT level. The DFT density of states (DOS) is then used to identify an appropriate ``correlated'' energy window that contains the desired orbitals.
For this selected energy window, the so-called Wannier projectors~\cite{le.ge.06,aichhorn2009,james_2021} are constructed, which are used to project the lattice Green's function
onto the localized Wannier-orbitals representation.
The resulting local Green's function serves as an input to the DMFT.
Here we employ the CT-QMC solver to obtain the self-consistent solution of the DMFT equations. 
The self-consistently obtained self-energy (with the double-counting removed) is then upfolded back to the Bloch basis so that it is suitable to update the lattice Green's function.

The charge density matrix is obtained from the lattice Green's function by summing over the Matsubara frequencies. This is then used to generate the total DFT+DMFT density matrix which is generally non-diagonal in the Kohn-Sham basis.
However, this total density matrix can be diagonalized into the orthonormal L\"{o}wdin-type basis~\cite{lowdin_quantum_1955} with a new set of diagonal DFT+DMFT occupations $N_{\zeta}^{\mathbf{k},\sigma} $ and wavefunctions ${\phi^\sigma_{\mathbf{k}, \zeta}(\mathbf{r})}$, as described in Ref.~\onlinecite{james_2021},
which are then used to update the electron density by 
\begin{equation}
    \rho(\mathbf{r})=\text{Tr}_\sigma \sum_{\zeta,\mathbf{k}} N_{\zeta}^{\mathbf{k},\sigma} 
\phi^\sigma_{\mathbf{k}, \zeta}(\mathbf{r})
\left(\phi^{\sigma}_{\mathbf{k}, \zeta}(\mathbf{r})\right)^\dagger.
\end{equation}
Here $\sigma$ is the spin index, and $\zeta$ is the eigenstate index.
The  DFT+DMFT results presented in this work are obtained by the fully charge self-consistent method with spin-polarized DFT inputs. 
Fully charge self-consistency is achieved by updating $\rho(\mathbf{r})$
from the DFT+DMFT occupations and wavefunctions. Subsequently, the Kohn-Sham equations are solved once, and the new Wannier projectors are generated for the next fully charge self-consistent cycle.

In the current implementation of the DMFT framework, the effective Anderson impurity problem corresponding to the correlated many-body system is solved by the CT-QMC method~\cite{gu.mi.11} using the TRIQS/CTHYB application~\cite{seth2016}. 
The CT-QMC methods have different formulations, namely the interaction expansion (CT-INT)~\cite{ru.sa.05}, the auxiliary-field (CT-AUX)~\cite{gu.we.08}, and the hybridization expansion (CT-HYB)~\cite{we.co.06}.
We use the CT-HYB formulation for all finite temperatures. CT-HYB operates on an imaginary time and frequency (Matsubara) axis. Therefore, analytical continuation is necessary to produce spectral functions on the real-frequency axis.
The latter suffers from the finite-precision arithmetic, which tends to amplify numerical noise, and produce unphysical artefacts. These issues may be especially severe for multi-orbital problems with complicated spectral lines.

The computation of the electron momentum density, $\rho(\mathbf{p})$, however, does not involve analytical continuation.
It is formally derived from the Fourier transformed real space wavefunctions $\psi^\sigma_{\mathbf{k}, \eta}(\mathbf{r})$, and in practice is determined by using the tetrahedron interpolation method for the discrete $\mathbf{k}$-mesh~\cite{Ernsting_2014}. The calculated electron momentum density has the form
\begin{equation}
  \rho^\sigma(\mathbf{p})= \sum_{\mathbf{k}, \eta}n^{\sigma}_{\mathbf{k}, \eta}\left|\int\limits_{V} \exp (-i \mathbf{p} \cdot \mathbf{r}) \psi^\sigma_{\mathbf{k}, \eta}(\mathbf{r}) \mathrm{d} \mathbf{r}\right|^{2}, \label{EMD}
\end{equation} 
where $n_{\mathbf{k}, \eta}$ are occupation functions with eigenstate index $\eta$. The electron momentum density within the DFT formalism is computed using the Kohn-Sham wavefunctions and occupation functions. Eq.~\eqref{EMD} can also be used in the DFT+DMFT formalism but now with the corresponding DFT+DMFT wavefunctions and occupation functions. By doing so, the direct impact of the non-physical artifacts of the analytical continuation on the electron momentum density can be circumvented. The occupations, as well as other observables, are implicitly dependent on the wavefunctions.
The changes in these quantities are coupled to those in the wavefunctions and are hard to disassociate.
Once the electron momentum density has been calculated, $J(p_{z})$ and/or $J_{\mathrm{mag}}(p_z)$ can be determined by Eqs.~\eqref{eq:J} and~\eqref{eq:J_mag}.

%%%%%%%%%%%%%%%%%%%%%%%%%%%%%%%%%%%%%%%%%%%%%%%%%%%%%%%%%%%%%%%%
\section{Computational Details and Results}
\label{sec:comp}

Both KKR and ELK self-consistent computations were performed with the same parameters for the crystal structure ($a=3.52$~\AA) and the same parametrization for the DFT exchange-correlation potential, LSDA~\cite{perdew1992}. 
The ELK DFT calculations used a 20$\times$20$\times$20 $k$-mesh, which proved to be sufficient for the k-point convergence of the self-consistent calculation. 
The KKR calculations within atomic sphere approximation  were performed on a 57$\times$57$\times$57 $k$-mesh, and a semicircular complex contour was used with $40$ energy points enclosing the one-particle poles of the Green's function.  
The minor differences in density of states and spectral functions can be attributed to the different approaches within ELK and KKR.

Sightly more significant differences are expected to appear at the DFT+DMFT level. Both approaches use a rotationally invariant form for the interacting Hamiltonian. The multi-orbital interaction has been parameterized by the average screened Coulomb interaction $U$ and the Hund's exchange coupling $J$. The values of $U$ and $J$ are sometimes used as fitting parameters, although recent developments allow the computation of the dynamic electron-electron interaction matrix elements exactly~\cite{ar.im.04}. 
It was shown \cite{mi.ar.08} that the static limit of the screened energy-dependent Coulomb interaction led to the $U$ parameter being in the energy range of $2$ and $4$~eV for all $3d$ transition metals.
Previous DMFT calculations showed that these $U$ and $J$ parameters provide the best description of the ground state properties related to the structure and different spectroscopic measurements for many of the $3d$ metals~\cite{ma.mi.09,ka.li.02,li.ka.01,ko.pa.12}.
In a considerable number of studies  of bulk fcc Ni the excellent agreement with the experimental results were obtained by setting ${J = 0.9}$~eV~\cite{ch.be.14a,ch.be.14b,be.mi.12,li.ka.01,ce.we.16}, the value which we also use here.
Besides, these $U$ and $J$ parameters are in line with constrained random-phase approximation (cRPA) calculations of $3d$ transition metals~\cite{mi.ar.08,sa.fr.11}.
Note, however, that the multi-orbital interacting Hamiltonian is formulated in different basis sets. In KKR+DMFT, the local atomic basis set is used~\cite{mi.ch.05,mi.eb.17}, and consequently, the many-body problem is formulated within the $d$-block. Correlation effects are felt by other orbitals only through the self-consistency cycle. 
In contrast, with the ELK+DMFT, the Wannier projectors are constructed such that the Ni-$d$ states, which are completely within the used correlated energy window of ${[-10, 3]}$~eV, are captured.
Further essential parameters for the CT-QMC computation~\cite{seth2016}) are the number of 4.2$\times$10$^{8}$ sweeps and the inverse temperature $\beta$ of $40$~eV$^{-1}$.
In both methods the spin-polarized around-mean-field double-counting term (AMF)~\cite{pe.ma.03,cz.sa.94} was employed.

The ELK+DMFT spectral function presented in Sec.~\ref{sec:sf} was calculated by analytically continuing the DMFT self-energy using the {\em LineFitAnalyzer} technique of the maximum entropy analytic continuation method implemented within the TRIQS/Maxent application \cite{maxent}.

The different descriptions of the potentials, full potential in ELK and the atomic sphere approximation (ASA) in KKR, also lead to the difference in the calculated chemical potentials. 
Within the KKR+DMFT method, 
the self-energy is added into the Kohn-Sham-Dirac equation~\cite{mi.ch.05,mi.eb.17}, and the chemical potential is updated to conserve the number of valence electrons similarly as in the DFT loop. 
The ELK+DMFT, using the Wannier projectors instead, updates the electron density from which a new set of Kohn-Sham eigenvalues and eigenvectors are generated and the corresponding chemical potential is obtained. The difference in the DFT+DMFT chemical potential with respect to the DFT values is at most a few tenths of an eV. The different solvers produce slightly different values for the real parts of the self-energies at the chemical potential. 
An important point here is the double counting and even though the functional form is the same for both KKR+DMFT and ELK+DMFT, the slightly different values in the occupation matrix produce slightly different double counting values.

%%%%%%%%%%%%%%%%%%%%%%%%%%%%%%%%%%%%%%%%%%%%%%%%%%%%%%%%%%%%%%%%
\subsection{$U$-dependent spin and orbital magnetic moments}
\label{sec:spin_mag}

To identify the optimal value of $U$, or at least to narrow the {\it ab-initio} interval, we first analyzed the behavior of the Ni ferromagnetic spin magnetic moment with respect to the on-site Coulomb interaction, $U$ and fixed Hund's rule coupling, ${J=0.9}$~eV.

%%%%%%%%%%%%%%%%%%%%%%%%%%%%%%%%%%%%%%%%%%%%%%%%%%%%%%%%%%%%%%%%
\begin{figure}[!t]%[t!]
  \centerline{\includegraphics[width=0.95\linewidth]{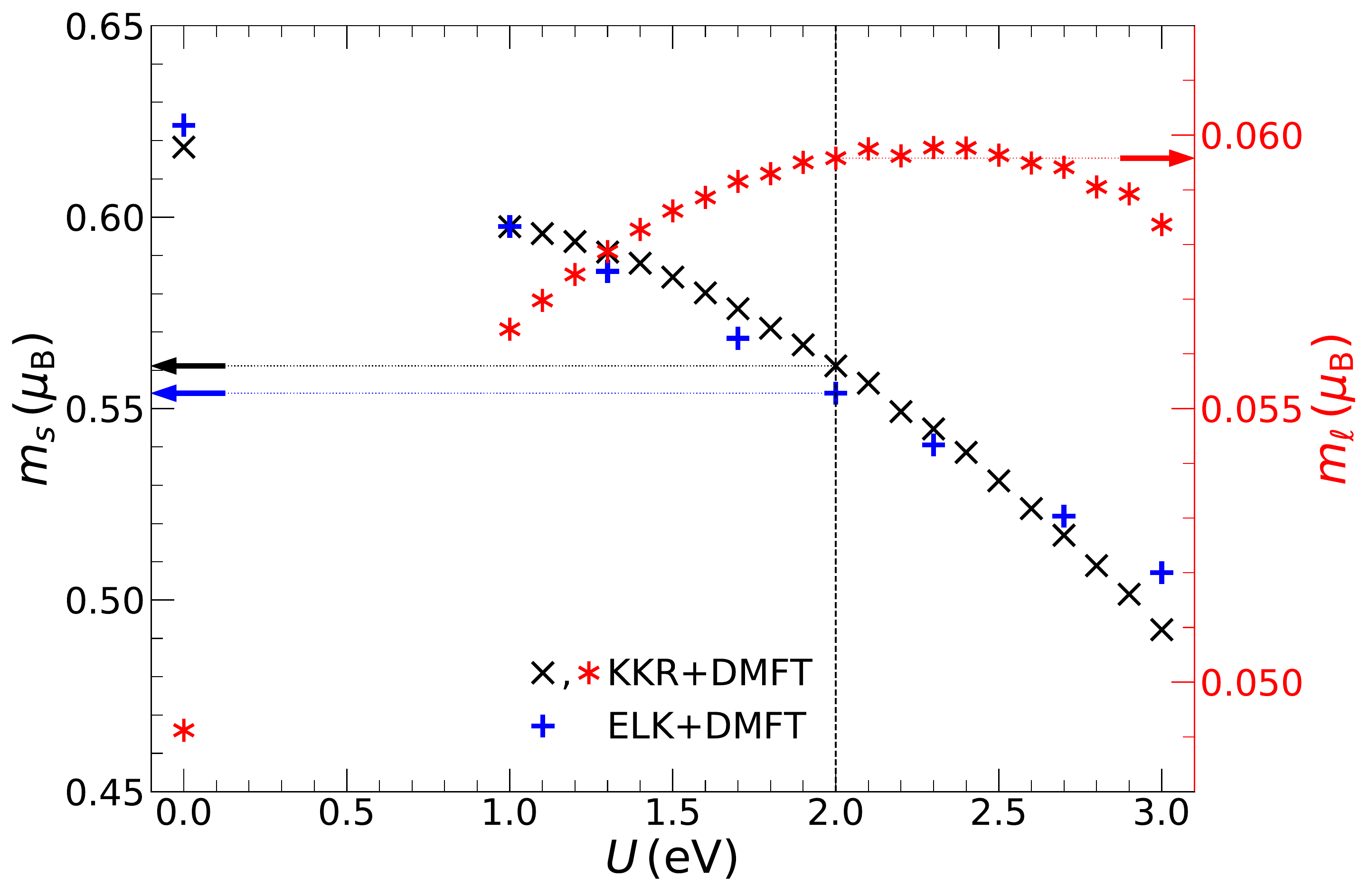}}
  \caption{Spin $m_s(\mu_B)$ and orbital $m_\ell(\mu_B)$ magnetic moment as function of the intra-site Coulomb potential $U$. Blue plus signs and black crosses represent results of ELK(+DMFT) and KKR(+DMFT) spin magnetic moment calculations, respectively. On the right axis we plot the KKR (red asterisks) orbital magnetic moment ($m_l$). In all calculations, the Hund's rule coupling parameter ${J=0.9}$~eV was used. The data points for ${U=0.0}$~eV represent the DFT calculations.}
 \label{fig:mag}
\end{figure}
%%%%%%%%%%%%%%%%%%%%%%%%%%%%%%%%%%%%%%%%%%%%%%%%%%%%%%%%%%%%%%%%

The ferromagnetic spin ($m_s$) and orbital ($m_\ell$) magnetic moment as a function of the on-site Coulomb interaction $U$ are shown in Fig.~\ref{fig:mag}. Both the ELK+DMFT and KKR+DMFT results show a similarly decreasing spin magnetic moment with increasing $U$, in quite close correspondence to each other. 

Contrary to the decreasing spin moment over the entire $U$ range, the orbital moment $m_\ell$ obtained in relativistic KKR+DMFT calculations, increases with the $U$ values, passing the maximum value at ${U\approx2.3}$~eV, and decreases upon further increasing the value of $U$. Even for the largest value of $U$ (${U=3.0}$~eV in the presented calculations), the KKR+DMFT orbital magnetic moment is larger than the corresponding DFT value by about $30\%$. Similar results have also been reported previously in Ref.~\onlinecite{mi.ma.14} and were interpreted as a correlation-induced orbital moment enhancement.

Despite the different descriptions, it is satisfying to see the good agreement between the results obtained with both methods. For ${U=2.0}$~eV, the calculated spin moment matches best with experiment {$\approx 0.56~\mu_{\mathrm{B}}$} for both DFT+DMFT methods and is within the {\it ab-initio} predictions for the $3d$ transition elements. These ${U=2.0}$~eV and ${J=0.9}$~eV values are in agreement with that used in the previous spin-polarized Ni ACAR study~\cite{ce.we.16}.
The experimental spin moment originates from the polarized neutron diffraction measurements by Ref.~\onlinecite{Mook_1966}. The total (spin + orbital) measured magnetic moment, which the analysis relied on, was subsequently revised by Ref.~\onlinecite{Danan_1968} to ${0.616\,\mu_B}$, and with which our KKR+DMFT ${U=2.0}$~eV calculation has excellent agreement.
Our chosen $U$ and $J$ values for ferromagnetic fcc Ni have a higher $J$/$U$ ratio compared to previous (DFT+)DMFT studies in the paramagnetic phase~\cite{ge.me.13}.
For other values of ${J}$ two different values of $U$ are required to match either spin or orbital moments, where as for ${J=0.9}$~eV and ${U=2.0}$~eV we obtain an excellent agreement of both. 
Magnetic Compton scattering, however, does not directly provide information concerning the orbital moments, but when combined with a SQUID measurement of the total magnetic moment, the orbital contribution can be inferred \cite{du.03}.

%%%%%%%%%%%%%%%%%%%%%%%%%%%%%%%%%%%%%%%%%%%%%%%%%%%%%%%%%%%%%%%%
\subsection{Magnetic Compton profiles}
\label{sec:mcp}

In the KKR(+DMFT), the magnetic Compton profiles are calculated from the spin-resolved momentum density ${\rho^{\sigma}(\mathbf{p})}$ which in turn is obtained as a contour integral of the Green's function in the momentum representation, Eq.~\ref{eq:rho_KKR}.  
In the ELK+DMFT the electron momentum densities (and the MCPs) are computed through the wavefunctions and occupation functions across the Brillouin zone on the imaginary frequency axis. 
The method of obtaining the wavefunctions and occupation functions in ELK-TRIQS are described in Ref.~\onlinecite{james_2021}.
In both methods, the MCPs were calculated within a sphere of radius ${16}$~a.u. (${|\mathbf p| \leqslant 16}$~a.u.), and then renormalized such that their areas were equal to the corresponding spin magnetic moment.

%%%%%%%%%%%%%%%%%%%%%%%%%%%%%%%%%%%%%%%%%%%%%%%%%%%%%%%%%%%%%%%%
\begin{figure*}[!t]
  \centerline{\includegraphics[width=\linewidth]{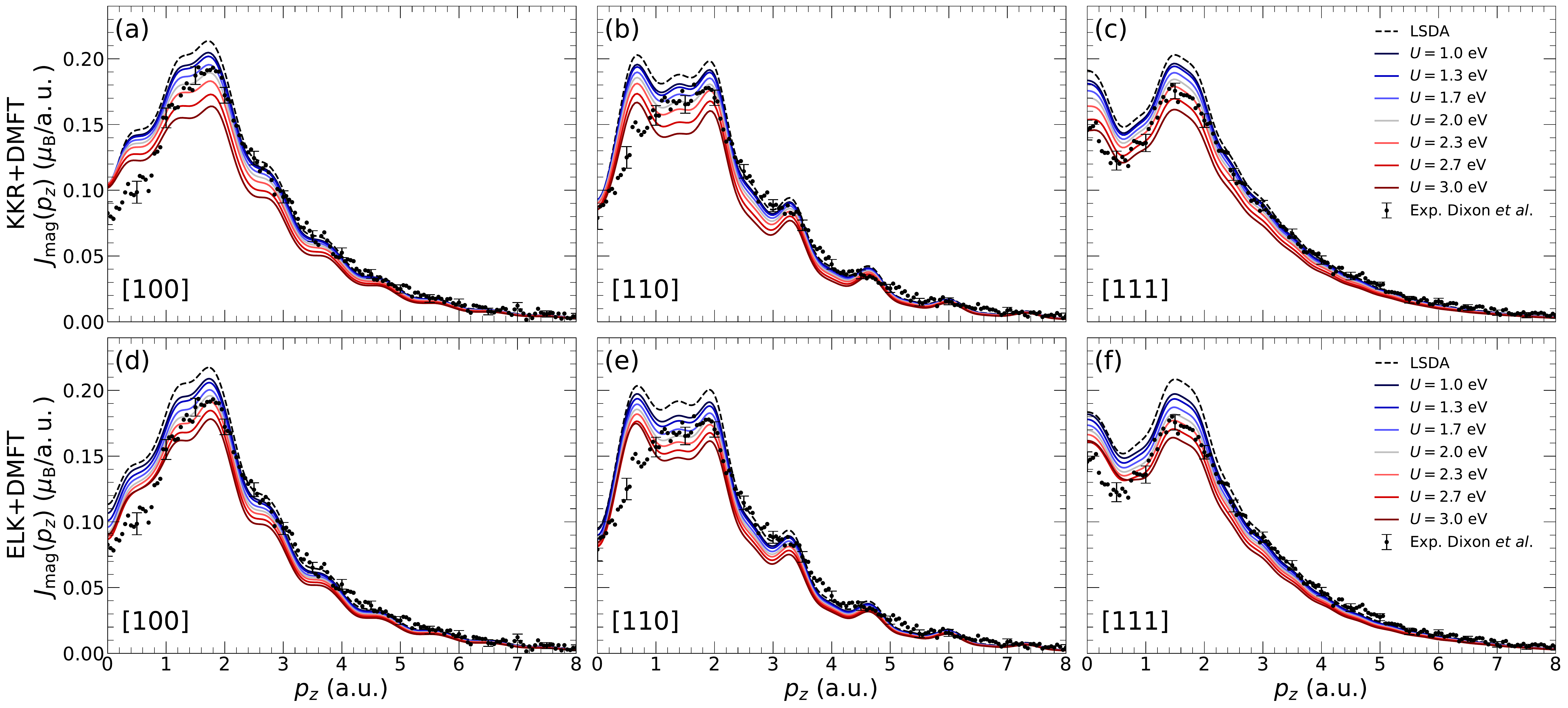}}
  \caption{The Ni magnetic Compton profiles (MCPs) at $[100]$, $[110]$, and $[111]$ high symmetry directions (indicated on each plot) for several intra-site Coulomb potential $U$ and fixed Hund's rule coupling ${J=0.9}$~eV. The KKR+DMFT MCPs results are shown in the (a), (b) and (c) panels (upper row). The ELK+DMFT MCPs are presented in the (d), (e) and (f) panels (lower row). The areas of each MCP have been normalized to their corresponding spin magnetic moment results given in Fig.~\ref{fig:mag}. The DFT+DMFT results are complemented by the LSDA results from the respective ELK and KKR codes (dashed curves) and the experimental measurements from Dixon {\it et~al.} (dots with error bars) \cite{di.du.98}. For clarity, the error bars are shown for every tenth data point. The computed results have been convoluted with a Gaussian with a full-width-at-half-maximum (FWHM) of $0.43$~a.u. to represent the experimental resolution.
}
 \label{fig:elk_kkr}
\end{figure*}
%%%%%%%%%%%%%%%%%%%%%%%%%%%%%%%%%%%%%%%%%%%%%%%%%%%%%%%%%%%%%%%%

%%%%%%%%%%%%%%%%%%%%%%%%%%%%%%%%%%%%%%%%%%%%%%%%%%%%%%%%%%%%%%%%
\begin{figure}[t!]
  \centerline{\includegraphics[width=0.95\linewidth]{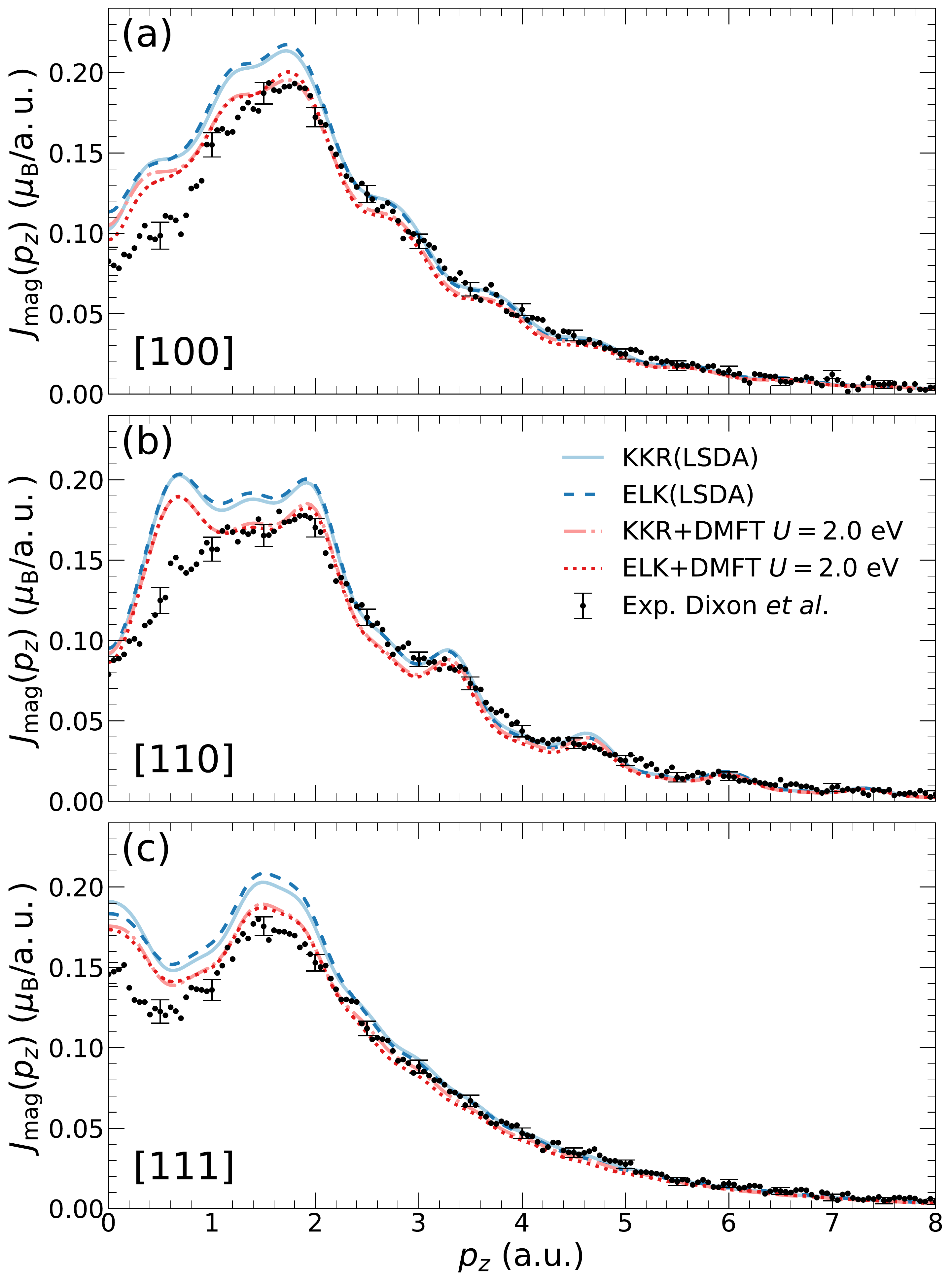}}
  \caption{The comparison of the experimental Ni magnetic Compton profiles (MCPs) from Dixon {\it et~al.}~\cite{di.du.98} (dots with error bars) with the DFT results (solid and dashed curves) and the DFT+DMFT results for the chosen ${U=2.0}$~eV and ${J=0.9}$~eV (dash-dotted and dotted curves). For clarity, the error bars are shown for every tenth data point. The (a), (b) and (c) panels show the MCPs for the $[100]$, $[110]$ and $[111]$ high symmetry directions. The computed results have been convoluted with a Gaussian with a full-width-at-half-maximum (FWHM) of $0.43$~a.u. to represent the experimental resolution.
  The areas of each MCP have been normalized to their corresponding spin magnetic moments given in Fig.~\ref{fig:mag}.}
 \label{fig:Exp_comp}
\end{figure}
%%%%%%%%%%%%%%%%%%%%%%%%%%%%%%%%%%%%%%%%%%%%%%%%%%%%%%%%%%%%%%%%

%%%%%%%%%%%%%%%%%%%%%%%%%%%%%%%%%%%%%%%%%%%%%%%%%%%%%%%%%%%%%%%%
\begin{figure}[t!]
  \centerline{\includegraphics[width=0.965\linewidth]{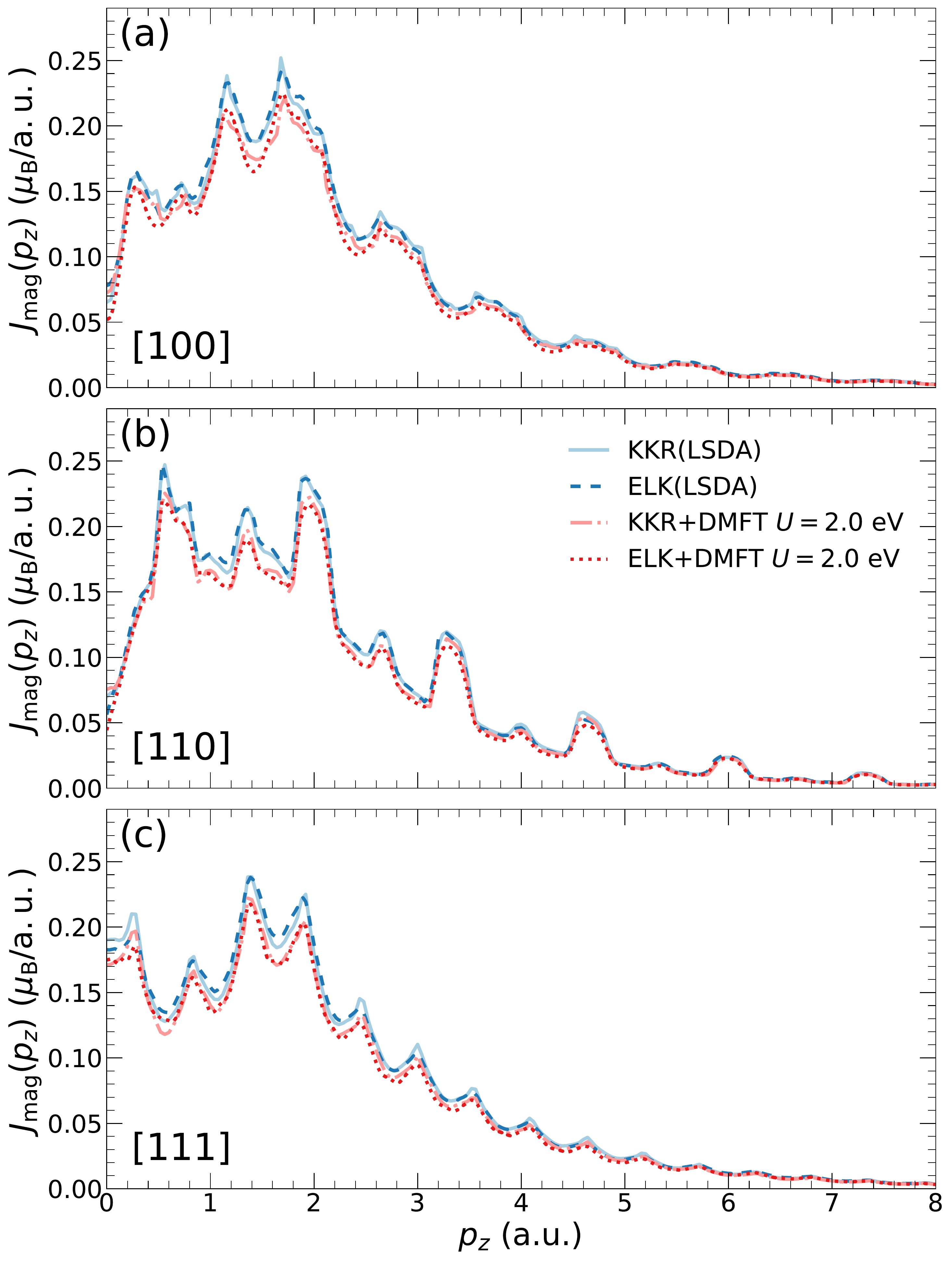}}
  \caption{The comparison of the unconvoluted high symmetry direction ([100], [110], and [111]) MCPs from the DFT (solid and dashed curves) and the DFT+DMFT with ${U=2.0}$~eV, ${J=0.9}$~eV (dash-dotted and dotted curves) calculations. The profiles in this figure are the unconvoluted counterparts of the MCPs which are in Fig.~\ref{fig:Exp_comp}}
 \label{fig:comp_unconv}
\end{figure}
%%%%%%%%%%%%%%%%%%%%%%%%%%%%%%%%%%%%%%%%%%%%%%%%%%%%%%%%%%%%%%%%

To analyze the effects of correlation on the MCPs, we calculated MCPs with the DFT+DMFT method for a series of on-site interaction values $U$ and Hund's rule coupling ${J=0.9}$~eV by employing both KKR+DMFT and ELK+DMFT. Fig.~\ref{fig:elk_kkr} shows the Ni MCPs along the cubic high symmetry directions, obtained using the KKR(+DMFT) (Figs.~\ref{fig:elk_kkr}~(a)-(c)),  and the ELK(+DMFT) (Figs.~\ref{fig:elk_kkr}~(d)-(f)) in the momentum range ${0 \leqslant p_z \leqslant 8}$~a.u..
The theoretical MCPs have been convoluted with a Gaussian with a full-width-at-half-maximum (FWHM) of $0.43$~a.u. to represent the experimental resolution.

\begin{figure*}[t!]
  \centerline{\includegraphics[width=0.95\linewidth]{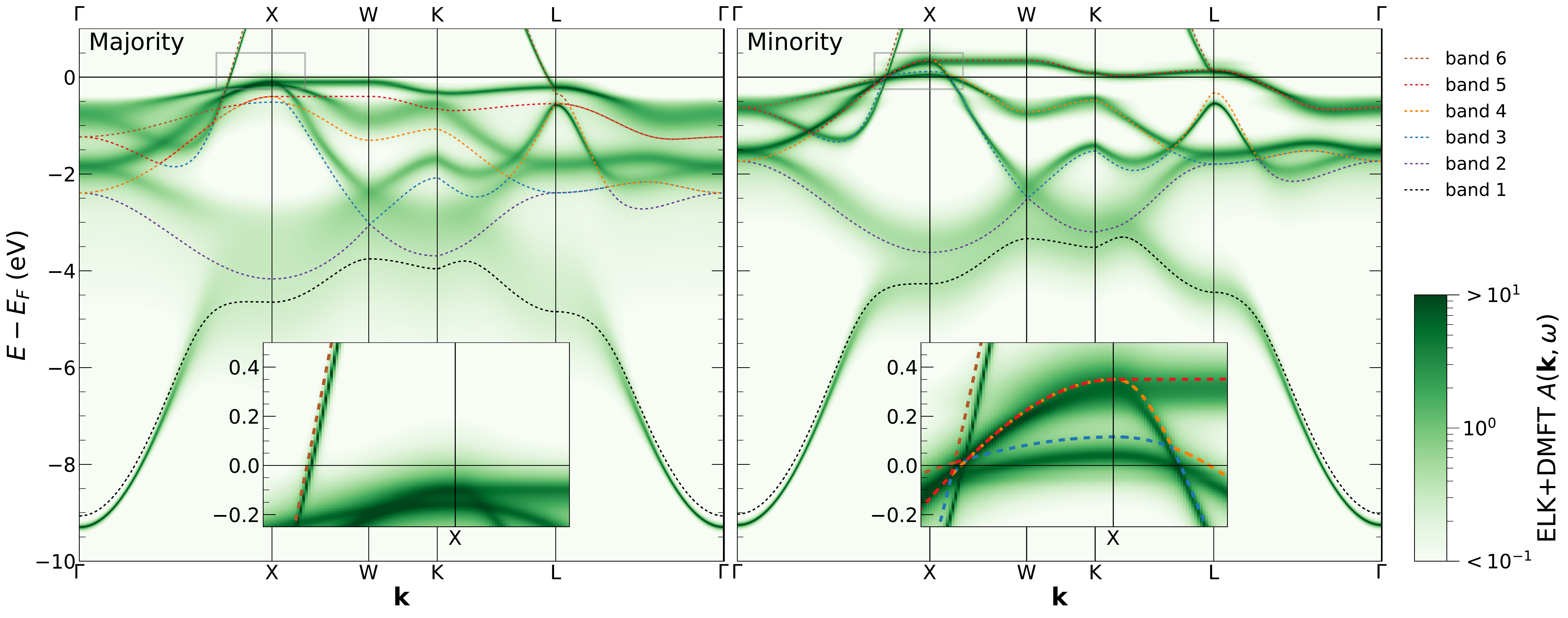}}
  \caption{The ELK DFT band structure and ELK+DMFT DFT+DMFT (${U=2.0}$~eV and ${J=0.9}$~eV) spectral function for the majority (left) and minority (right) spin. The DFT bands have been broken down in terms of their indices (using the same numbering as Dixon {\it et~al.}~\cite{di.du.98}) for discussions about the band contributions to the MCPs and resemble those of Dixon {\it et~al.}. The insets are a zoomed image around the X symmetry point (indicated by the grey outline) showing the differences between the theoretical treatments.}
 \label{fig:Akw}
\end{figure*}

Starting with the presented DFT results, the MCPs show good agreement with the experiment for ${p_z > 2}$~a.u. but these MCPs do not match the low-momentum region for any of these high symmetry directions. Our DFT results are in good agreement with those previously presented in Ref.~\onlinecite{di.du.98}.
The MCP peak structures within the first Brillouin zone are due to the exchange splitting, which in turn causes the majority and minority spin bands to cross the Fermi level at different $\mathbf{k}_F$ values (see Fig.~\ref{fig:comp_unconv} and Fig.~\ref{fig:Akw}).
These peaks are periodically repeated in the MCPs as these are the umklapp contributions from higher zones (i.e., ${\mathbf{k}+\mathbf{G}}$ where $\mathbf{G}$ is the reciprocal lattice vector). 
One of the advantages of the effective one-particle framework of DFT calculations is the possibility to decompose the total MCP into the contributions originating from individual bands~\cite{di.du.98,ku.as.90,ti.br.90,ka.ku.03}.
The dip in this low-momentum region has been attributed partly to the contribution of the so-called negative polarization of the $s$- and $p$-like bands with respect to the positive contribution of the $d$-bands.
At the same time, Refs.~\onlinecite{di.du.98,ka.ku.03} note that another source of discrepancy may be due to  the $d$-like fifth band (band numbering according to Ref.~\onlinecite{di.du.98}), where Ref.~\onlinecite{ka.ku.03} attributes the shape of the contribution of this band to the inconsistencies between the theoretical and the true Fermi surface.
These interpretations, based on the DFT band structures, raise some interesting unsolved questions about the origin of the discrepancy at low momentum. 
From the DFT results, the predicted negative polarization contributions are not sufficient to explain the low momentum dip seen in the experimental results. 
Dixon {\it et~al.}, Ref.~\onlinecite{di.du.98}, suggested that it was the deficient representation of the $d$-electron correlations in LSDA (and GGA), not just the negative polarization from the $s$- and $p$-electrons, which was the potential cause for the low momentum experiment-theory disagreement ~\cite{di.du.98}. Artificial shifts of the bands around the Fermi level~\cite{ma.du.04} showed improved agreement with the low momentum MCP region. As correlation effects lead to the shift of those bands naturally, improved theoretical description of the Ni MCPs can be obtained by taking them into account. 
Recent studies~\cite{be.mi.12,ch.be.14a,ch.be.14b} also demonstrated that including the local correlations through the DMFT framework reduces the discrepancy between theoretical and experimental MCPs of Ni.

Moving onto the DFT+DMFT results, 
the large dips near ${p_z= 0}$~a.u. in the high symmetry directions are better reproduced by the DFT+DMFT MCP for ${U>2.3}$~eV. On the other hand, for high-momentum, ${p_z>2}$~a.u., region ${U<2.3}$~eV is a better choice.
Although we are able to produce improved agreement (with respect to the DFT MCPs) with the experiment at low momentum, ${p_z < 1}$~a.u., DFT+DMFT fails to reproduce the experimental MCP for the $[100]$ and $[110]$ directions (see Fig.~\ref{fig:elk_kkr}~(a),(d) and (b),(e)). Along the nearest-neighbor direction $[110]$, no $U$ value was found to suppress the peak at around ${p_z = 0.6}$~a.u..
Although the general low momentum disagreement is the case for both implementations, there are some notable differences between the ELK+DMFT and KKR+DMFT results. Along $[100]$ direction, ${J_{\mathrm{mag}}(p_z)}$ for ${p_z<1}$~a.u calculated with ELK+DMFT for increasing values of $U$ matches the experimental MCP better than those obtained with KKR+DMFT. The latter visibly overestimates the ${J_{\mathrm{mag}}(p_z)}$ (by almost the same amount) for all $U$ values considered.
The opposite happens for the $[111]$ direction. In this case, ${J_{\mathrm{mag}}(p_z)}$ obtained with KKR+DMFT matches the experimental values in ${p_z\lesssim 2}$~a.u. region for ${U\geqslant 2.3}$~eV, while ELK+DMFT results overestimate the experimental values for ${p_z\lesssim  1}$~a.u for all considered values of $U$.

Although in general the low momentum is better described with higher $U$ values (see Fig.~\ref{fig:elk_kkr}), the costs of this is the poorer agreement with the experiment from $1$~a.u. to about $5$~a.u..
This is because the area under the MCP, which is equal to the corresponding spin moment for each $U$ value in Fig.~\ref{fig:mag}, reduces with increasing $U$ and is less than the experimental value for about ${U>2.0}$~eV. Therefore, for the different $U$ values, an improvement in one momentum region of the MCP causes another region to worsen in order to conserve the area. 

We did not find a single $U$ value, within the {\it ab-initio} range of $U$ values, which would simultaneously match both, low- and high-momentum regions of experimental profile within its error.
On the other hand, in the previous section, we identified that the DFT+DMFT calculation with ${U=2.0}$~eV and ${J=0.9}$~eV produces the best match between the calculated and experimental magnetic moment.
To see how well the DFT+DMFT MCPs for ${U=2.0}$~eV match the experimental MCPs from Ref.~\onlinecite{di.du.98}, and also to compare the results obtained by two different packages
and two distinct frameworks
in Fig.~\ref{fig:Exp_comp}, we show the corresponding MCPs. 
Although the MCPs calculated in the DFT+DMFT framework for ${U=2.0}$~eV deviate from the experimental results in the momentum range ${0< p_z < 1}$~a.u., extending DFT with the DMFT framework significantly improves the description of the experiment in the range ${1< p_z< 2}$~a.u.. 
For ${U=2.0}$~eV, the structure of the MCPs is well reproduced in all three high symmetry directions in this region where the dominant contributions are made.
DFT+DMFT results also stay in reasonably good agreement with the experiment for higher values of $p_z$, from ${p_z=2}$~a.u. onwards, but they tend to slightly underestimate the tails, although they are within the experimental error. This is also a consequence of the calculations overestimating the low momentum region.

%%%%%%%%%%%%%%%%%%%%%%%%%%%%%%%%%%%%%%%%%%%%%%%%%%%%%%%%%%%%%%%%
%%%%%%%%%%%%%%%%%%%%%%%%%%%%%%%%%%%%%%%%%%%%%%%%%%%%%%%%%%%%%%%%

Overall, dynamic correlations improve the agreement with the experimental data beyond the LSDA results.
The results including dynamic correlations also show the correct trend for low momentum region ${p_z \lesssim 2}$~a.u. where better MCPs are obtained in comparison to the LSDA.
LSDA overestimates MCP values almost in the entire region.
As mentioned earlier, since the areas under the MCPs directly equal the spin moment ($m_s$), the areas reduce with increasing $U$ as per Fig.~\ref{fig:mag}.
Nevertheless, since DFT+DMFT also overestimates the experimental MCP values in the ${p_z \lesssim 1}$~a.u. region, the consequence is that the tails (high-momentum region) are underestimated --- after all, the areas beneath the MCP and ${U=2.0}$~eV DFT+DMFT curves are almost equal.
The current results, however, do not allow us to infer the optimal value for the on-site Coulomb interaction necessary to obtain the best agreement with the experimental measurements. Nevertheless, we see that $U$ values in the range $[1.7,2.3]$~eV describe the on-site Coulomb interaction reasonably well (almost exactly within experiment error bars in the ${1\lesssim p_z \lesssim 2}$~a.u. range), in agreement with positron annihilation measurements~\cite{ce.we.16}. Similar conclusions have been drawn in previous papers reporting the correlation effects upon the MCP of Ni~\cite{be.mi.12,ch.be.14a,ch.be.14b}.

A direct comparison between the methods can be seen in Fig.~\ref{fig:comp_unconv}, where we plot the theoretical MCPs which have not been convoluted with the experimental resolution.
The results produced with the two DFT+DMFT implementations are in excellent agreement. Therefore, we are confident that the effect of the resolution on the MCPs does not hide any glaring disagreements between the implementations.

Finally, we conclude that neither implementation produces results (for all $U$ values) that have a better overall agreement with the experimental data than the other (within the experimental error). The level of experiment-theory agreement between the MCPs from both implementations varies in different regions of momentum. 
Overall, the results from the two implementations are in good agreement with minor discrepancies due to the aforementioned differences discussed at the beginning of Sec.~\ref{sec:comp}. 

%%%%%%%%%%%%%%%%%%%%%%%%%%%%%%%%%%%%%%%%%%%%%%%%%%%%%%%%%%%%%%%%
\subsection{Spectral function}
\label{sec:sf}

Features in the MCPs can be traced back to the form of the spectral function which, for the non-interacting case, is represented by the band structure. 
Fig.~\ref{fig:Akw} shows the DFT band structure together with the DFT+DMFT $\mathbf{k}$-resolved spectral function along the high symmetry direction in the Brillouin zone from the ELK and ELK+DMFT calculations.  
In the present DFT calculation, we confirm that the bands $5$ and $6$ of Fig.~\ref{fig:Akw} only give a positive contribution to the MCPs whereas bands $1$-$4$ have negative contributions to the MCPs at low momentum. 

In a many-body picture, however, such a band-resolved  interpretation is not possible. The spectral function in Fig.~\ref{fig:Akw} shows the quasi-particle dispersion. The self-energy affects the two spin channel spectral functions differently. A significant part of the energy dependence of the self-energy is related to the different occupations of the spin-polarized $d$-states, on which the MCPs are also dependent. Scattering processes involving $s$-electrons may be neglected as the corresponding orbitals are almost completely filled~\cite{lieb.79}.

Within the DMFT approximation, the self-energy matrix is diagonal in the angular momentum representation and is independent of ${\mathbf{k}}$.
It is the orbital dependence of the self-energy that produces a coupling between the terms of the $d^8$-multiplets
~\cite{lieb.79}, where the neglected $k$-dependence of the self-energy amounts to disregarding the hopping processes of the two holes bound to the same Ni-site. The CT-HYB impurity solver captures the self-energy contributions relevant for the strong ferromagnetic state such as repeated scattering of paired holes, hole-hole and hole-electron interactions as these processes enter in the fully rotationally invariant formulation of the Hubbard model and parameterized by the $U$ and $J$ parameters~\cite{ko.pa.12}. As Ni has a relatively large band width, the atomic multiplet structure is extended in the energy range around ${-6}$~eV. Therefore, the expected satellite in our treatment is a broad feature instead. 
The prominent correlation effect of the DFT+DMFT $\mathbf{k}$-resolved spectral function is to renormalize the position and width of the $d$-bands and significantly reduce the exchange splitting to about $0.3$~eV at the L-point (which we measured as the difference between the majority and minority band centers). These are direct consequences of the presence of the real part of the DMFT self-energy having a negative slope at $E_F$.
These features are in good agreement with experiments \cite{hi.kn.79,ea.hi.80} and are in line with previous studies \cite{li.ka.01,br.mi.06,hausoel2017local}. 
We observe that the crossing of the bands at the Fermi level hardly changes for the majority spin channel (see the left panel in Fig.~\ref{fig:Akw}).
In the minority bands, however, there are subtle changes around the X-point, where two X-hole pockets reside (see the inset in the right panel in Fig.~\ref{fig:Akw}).
These changes are less significant for the MCPs but are relevant in other experiments such as de Haas-van Alphen and ARPES~\cite{Wang_1974,weling_1982}. Previous experiment-theory comparisons~\cite{Wang_1974,weling_1982} have shown that DFT predicts a second shallow minority hole pocket around X. This is referred to as the minority X$_2$ hole pocket (related to minority band $3$ in Fig.~\ref{fig:Akw}) but there is no strong evidence of its presence in the experiments.
The present DFT+DMFT calculation with ${U=2.0}$~eV shows that the size of the minority X$_2$ hole pocket shrinks and also becomes shallower as compared to DFT results, but it does not vanish.
These may indicate that other correlation effects are required to suppress this band below the Fermi level, or that the large effective mass of X$_2$ hole pocket due to the shallowness of the corresponding band around X (see the inset in the right panel in Fig.~\ref{fig:Akw}) might have made its observation more challenging. 

Contrary to previous interpretations based on the one-particle description it is not obvious that the negative polarization contributions to the MCP (by the $s$- and $p$-electrons) is the cause for the disagreement between the experiment and the DFT and DFT+DMFT computations.
The low momentum disagreement is likely the consequence of the other missing correlation effects beyond DFT+DMFT, such as screening.
As screening is a genuine many-body effect, it requires methods like quasiparticle GW (QSGW).
Such a calculation for Ni has been performed recently by  L. Sponza {\it et~al.}~\cite{sponza_2017}. The QSGW calculations produce an enhanced value for the magnetic moment and exchange splittings. 
Nonetheless, in supplementing the computations with DMFT in the combined QSGW+DMFT, the values for the magnetic moment and exchange splitting are in good agreement with the experiment. 
We expect that a QSGW+DMFT calculations would likely improve the MCPs, as these incorporate non-local and screening effects. 

%%%%%%%%%%%%%%%%%%%%%%%%%%%%%%%%%%%%%%%%%%%%%%%%%%%%%%%%%%%%%%%%
\section{Conclusion and Outlook}

To conclude, we have presented results of two different DFT+DMFT implementations to calculate the spin-resolved momentum distributions ${\rho^\sigma({\bf p})}$, and the magnetic Compton profile ${J_\mathrm{mag}(p_z)}$. 
Both of these implementations show excellent agreement with each other considering the differences in their approaches to applying both DFT and DMFT and the different challenges that these contribute to the calculations.

The DFT+DMFT spin moment calculations have the same $U$ dependence in both setups, the slight difference  in magnitude likely being due to the details of the implementations.
Although the spin moment improves to be comparable with the experimental value, the shape of the MCP has a weak $U$ dependence, features in the profile such as umklapp peaks, remain relatively unchanged and only the MCP contributions are redistributed compared to the calculated DFT profiles.
 
For the ${U=2.0}$~eV calculation, which reproduces the experimental spin (and total) magnetic moments, the corresponding spectral function reveals that the minority X$_2$ pocket shrinks and gets shallower with respect to the DFT calculations, but nevertheless still survives. This small X$_2$ pocket is likely to have a large effective mass and this may explain why it was not observed in the de Haas-van Alphen experiment.

According to our combined DFT+DMFT approaches, some arguments in previous DFT studies built upon the existence of negative polarization description are not sufficient to explain the discrepancy between the theoretical and experimental MCP and low-momentum region.
Instead, theories including a non-local description of interaction and retardation effects (i.e., energy-dependent screening) such as cluster-DMFT, GW (QSGW) and beyond might be more suitable to deliver a better description of the MCP in ferromagnetic metals such as Ni.
To truly resolve the intricacies which may arise between the aforementioned theoretical frameworks, it would be essential to remeasure the Ni MCPs with a higher resolution. This will lead to further valuable understanding of the many-body groundstate properties probed in momentum space.

\begin{acknowledgments}
Financial support by the Deutsche Forschungsgemeinschaft through TRR80 (project E2)
Project number 107745057 is gratefully acknowledged.
A.D.N.J.~acknowledges funding and support from the
Engineering and Physical Sciences Research Council (EPSRC), Grant No.~EP/L015544/1.
M.S.~was partially supported by the Shota Rustaveli Georgian National Science Foundation through the grant N~FR-19-11872.
We would like to thank E.~I.~Harris-Lee for his insights and also J.~Min\`ar and H.~Ebert for a fruitful collaboration.
\end{acknowledgments}

\end{document}